\documentclass[sigconf]{acmart} 
\usepackage{csquotes}
\usepackage{amsfonts,amsmath,amsthm}

\usepackage{float} 
\usepackage{lipsum}
\usepackage{multirow} 
\usepackage{tikz}
\usepackage{pgfplots}
\pgfplotsset{compat=1.18}
\usepackage{array}
\usepackage{enumitem}

\usepackage{makecell}
\AtBeginDocument{%
  }

\setcopyright{acmlicensed}
\copyrightyear{2025}
\acmYear{2025}
\acmConference[Preprint]{Preprint}{2025}{Preprint}
\begin{document}

\title{Semantic Search At LinkedIn}


\author{Fedor Borisyuk}
\author{Sriram Vasudevan}
\author{Muchen Wu}
\author{Guoyao Li}
\authornote{Work done while at LinkedIn.}
\author{Benjamin Le}
\author{Shaobo Zhang}
\author{Qianqi Kay Shen}
\author{Yuchin Juan}
\author{Kayhan Behdin}
\author{Liming Dong}
\author{Kaixu Yang}
\author{Shusen Jing}
\author{Ravi Pothamsetty}
\author{Rajat Arora}
\author{Sophie Yanying Sheng}
\author{Vitaly Abdrashitov}
\author{Yang Zhao}
\author{Lin Su}
\author{Xiaoqing Wang}
\affiliation{
 \institution{LinkedIn}
 \city{Mountain View}
 \state{CA}
 \country{USA}}
\email{fedorvb@gmail.com}

\author{Chujie Zheng}
\author{Sarang Metkar}
\author{Rupesh Gupta}
\author{Igor Lapchuk}
\author{David N. Racca}
\author{Madhumitha Mohan}
\author{Yanbo Li}
\author{Haojun Li}
\author{Saloni Gandhi}
\author{Xueying Lu}
\author{Chetan Bhole}
\author{Ali Hooshmand}
\author{Xin Yang}
\author{\mbox{Raghavan Muthuregunathan}}
\author{Jiajun Zhang}
\author{Mathew Teoh}\authornotemark[1]
\author{Adam Coler}
\author{Abhinav Gupta}
\affiliation{
 \institution{LinkedIn}
 \city{Mountain View}
 \state{CA}
 \country{USA}}
\email{razheng@linkedin.com}

\author{Xiaojing Ma}
\author{\mbox{Sundara Raman Ramachandran}}
\author{Morteza Ramezani}
\author{Yubo Wang}\authornotemark[1]
\author{Lijuan Zhang}
\author{Richard Li}
\author{Jian Sheng}
\author{Chanh Nguyen}
\author{Yen-Chi Chen}
\author{Chuanrui Zhu}
\author{Claire Zhang}
\author{Jiahao Xu}
\author{Deepti Kulkarni}
\author{Qing Lan}
\author{Arvind Subramaniam}
\author{Ata Fatahibaarzi}
\author{Steven Shimizu}
\author{Yanning Chen}\authornotemark[1]
\affiliation{
 \institution{LinkedIn}
 \city{Mountain View}
 \state{CA}
 \country{USA}}
\email{xjma@linkedin.com}

\author{Zhipeng Wang}
\author{Ran He}
\author{Zhengze Zhou}
\author{Qingquan Song}\authornotemark[1]
\author{Yun Dai}\authornotemark[1]
\author{Caleb Johnson}
\author{Ping Liu}
\author{\mbox{Shaghayegh Gharghabi}}\authornotemark[1]
\author{\mbox{Gokulraj Mohanasundaram}}
\author{Juan Bottaro}
\author{Santhosh Sachindran}
\author{Qi Guo}\authornotemark[1]
\author{Yunxiang Ren}
\author{Chengming Jiang}
\author{Di Mo}
\author{Luke Simon}\authornotemark[1]
\author{Jianqiang Shen}
\author{Jingwei Wu}
\author{Wenjing Zhang}
\affiliation{
 \institution{LinkedIn}
 \city{Mountain View}
 \state{CA}
 \country{USA}}
\email{zhipwang@linkedin.com}

\makeatletter

  \makeatother


\renewcommand{\shortauthors}{Fedor Borisyuk et al.}

\begin{abstract}
Semantic search with large language models (LLMs) enables retrieval by meaning rather than keyword overlap, but scaling it requires major inference efficiency advances. We present LinkedIn’s LLM-based semantic search framework for AI Job Search and AI People Search, combining an LLM relevance judge, embedding-based retrieval, and a compact Small Language Model trained via multi-teacher distillation to jointly optimize relevance and engagement. A prefill-oriented inference architecture co-designed with model pruning, context compression, and text–embedding hybrid interactions boosts ranking throughput by over 75× under fixed latency constraint while preserving near–teacher-level NDCG, enabling one of the first production LLM-based ranking systems with efficiency comparable to traditional approaches and delivering significant gains in quality and user engagement.
\end{abstract}



\maketitle

\renewcommand{\thefootnote}{\fnsymbol{footnote}}


\section{Introduction}
Search is central to LinkedIn’s mission of connecting professionals to economic opportunity.
Members rely on search to discover jobs, people, and knowledge, with growing expectations for
semantic relevance and personalization. Keyword-based retrieval struggles to capture natural
language intent, motivating a re-architecture of LinkedIn Search around large language models
(LLMs) that enable semantic understanding at LinkedIn scale. This paper presents the models,
training methodology, and system design that make LLM-based semantic search feasible at
hundreds of thousands of queries per second.

LinkedIn Search serves multiple high-impact verticals, most notably People Search and Job
Search. While they share a semantic foundation, they differ in intent, constraints, and user
interactions. People Search focuses on discovering member profiles via free-form queries and
structured filters (e.g., connection degree, location, company), supporting actions such as
profile views, connects, and messages. Job Search targets role discovery under constraints
(e.g., location, work modality, seniority), enabling iterative exploration from shortlisting
to application. Both verticals rely on shared semantic infrastructure that must scale reliably
while optimizing for relevance, personalization, and engagement.

Semantic search retrieves and ranks results by meaning rather than token overlap. At LinkedIn,
query execution begins with a \emph{Query Understanding (QU)} layer that converts
short, ambiguous queries into deterministic, machine-interpretable signals, including routing
decisions, normalized attributes, and query reformulations. These signals define a stable
semantic contract for all downstream stages across Semantic Job Search and Semantic People
Search~\cite{liu2025powering_job_search_scale}. Conditioned on QU outputs, the retrieval stack
embeds queries and candidates, performs large-scale vector retrieval, and produces a
high-recall candidate set, which is then reranked by higher-capacity LLM-based cross-encoders
followed by business logic and policy enforcement. The central challenge throughout this
pipeline is accurate relevance estimation under strict latency constraints.

High-quality semantic search also requires principled governance of what ``relevance'' means
in product context. We address this with SAGE~\cite{illuminator2025}, an LLM-based evaluation
framework that operationalizes relevance policy across model development, experimentation,
and launch decisions for Job Search and People Search. SAGE combines explicit product policy,
curated human-labeled precedent data, LLM surrogate judges, and simulation-driven iteration.
To meet scale constraints, frontier LLM judges are distilled into an in-house decoder-only 8B
student that outputs graded (0-4) relevance scores with natural-language rationales, achieving
strong alignment with human precedent (linear kappa 0.77) and its teacher (0.81) while
supporting tens of millions of evaluations per day. Within the search stack, SAGE supervises
ranking (\S\ref{SLM_ranking}) and retrieval
(\S\ref{subsection:semantic-search-retrieval}) and governs launch decisions.

Deploying LLMs in search systems at tens of thousands of QPS remains a major industrial
challenge. Although LLM cross-encoders substantially improve query-candidate interaction
modeling, their inference cost scales with context length, making long candidate text and
prefill-heavy computation the dominant production bottlenecks. Prior work demonstrates LLM
gains in semantic understanding, relevance estimation, and evaluation
\cite{pinterest2024searchllm,pinterest2025llmassess}, but many approaches focus on high-capacity
rerankers without addressing production-scale serving constraints
\cite{qwenreranker_paper,bytedance2025listconranker,lee2025geminiembeddinggeneralizableembeddings},
or apply LLMs only as auxiliary components within traditional pipelines
\cite{microsoft2025webscalellmrec,chen2024hllmenhancingsequentialrecommendations}. As a result,
compact encoder-based models remain dominant for real-time ranking
\cite{tencent2024transformer}. At retrieval time, most industrial systems rely on ANN methods
with recall-efficiency trade-offs, whereas our work builds on a GPU-accelerated exhaustive
retrieval stack deployed at LinkedIn scale
\cite{borisyuk2024linrmodelbasedneural,yuchin10.1145/3705328.3748116}, enabling full scans over
billion-scale indices with rich attribute-based filtering and without ANN
“liquidity” issues.

\paragraph{Key Contributions.}
\begin{itemize}
    \item \textbf{End-to-end semantic search at production scale.} We present a unified stack combining GPU-accelerated exhaustive retrieval, LLM-supervised embedding and retrieval-as-ranking models, and an SLM reranker powering LinkedIn Job and People Search under strict latency and throughput constraints.
    \item \textbf{Unified ranking via multi-teacher, multi-task distillation.} We introduce a
    distillation framework that transfers knowledge from specialized relevance and engagement
    teachers into a single student ranker using distributional supervision, relevance-aware
    warm starts, and imbalance-aware loss masking.
    \item \textbf{High-QPS LLM ranking via model-infrastructure co-design.} We enable practical
    LLM ranking through structured pruning, offline summarization for context compression,
    hybrid text--embedding interactions, and a scoring-optimized inference stack with
    prefill-only execution and shared-prefix amortization.\footnote{The scoring-specialized
    LLM inference stack has been open-sourced as part of the \texttt{sglang} project:
    \url{https://github.com/sgl-project/sglang}.}
\end{itemize}

\section{Modeling and Infra Innovations}\label{sec:arch}
Our system uses a two-stage architecture: a GPU-accelerated exhaustive embedding-based retriever generates candidates, followed by an SLM reranking the top 250 results in production. The SLM jointly predicts relevance and engagement and outperforms a DLRM-style baseline \cite{LiRank_paper}. We describe retrieval training, relevance-first and joint relevance-engagement optimization for the SLM, and pruning/serving/training techniques that make LLM ranking efficient at scale.

\subsection{SLM Ranking}\label{SLM_ranking}

\begin{figure}
    \centering
    \includegraphics[width=1.0\linewidth]{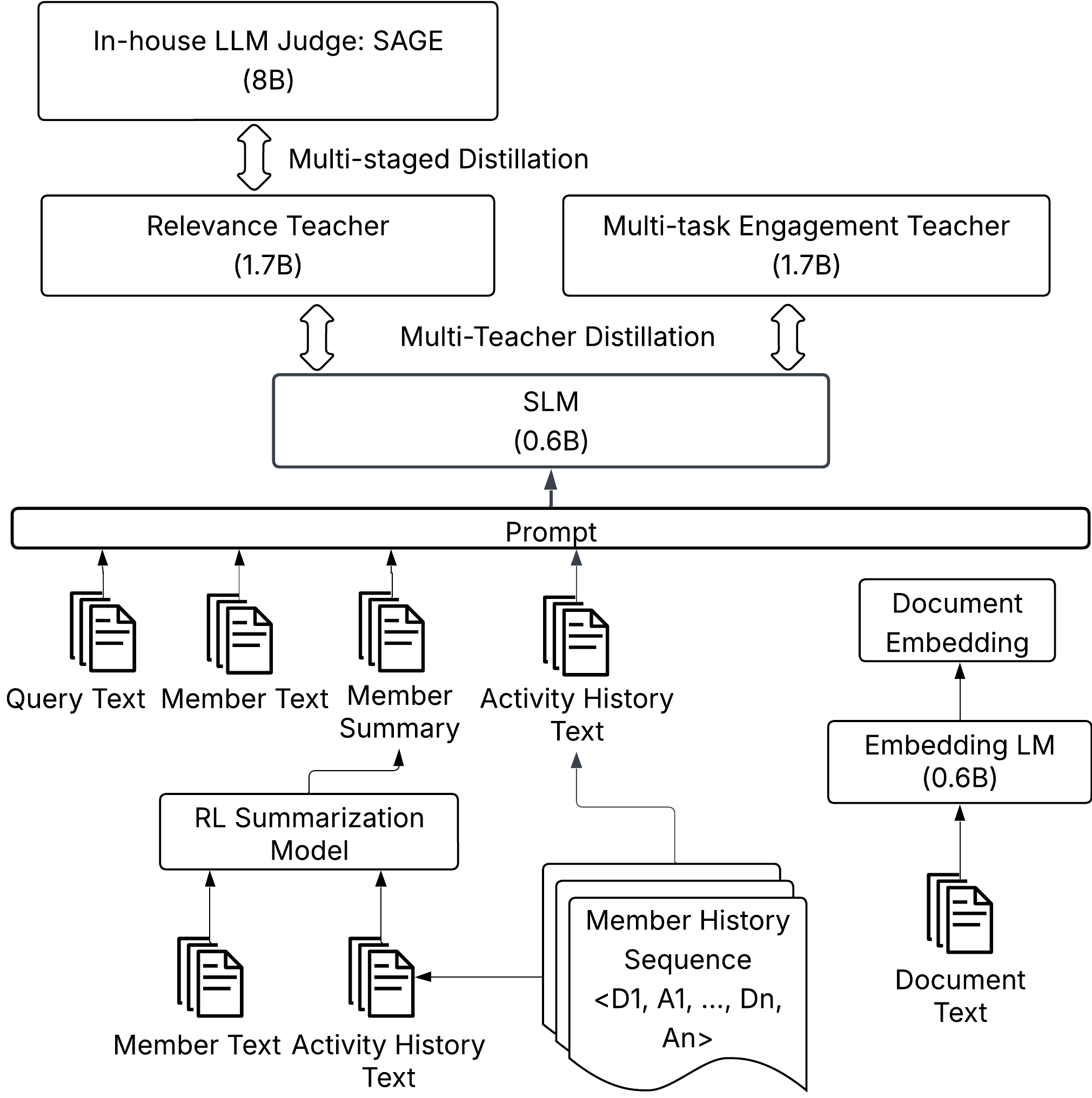}
    \captionsetup{font=small}
    \caption{An overview of the multi-stage training framework for SLM ranking in semantic search.}
    \vspace{-2mm}
    \label{fig:slm_ranking_overall}
\end{figure}

The ranking layer is the final heavy model before returning results. A unified SLM predicts multiple utility scores to balance relevance, engagement, and marketplace health.

LLMs are strong rankers due to broad language understanding, with evidence across preference extraction \citep{relevance4}, cold-start recommendation \citep{related1,related5}, and action prediction \citep{related3,related4}. We refer to~\citep{llmrs} for a detailed review of LLM-based search and recommender systems. In semantic search, LLMs  also show strong results \citep{sachan2022improving,zhuang2024setwise,zhang2024two,rankvicuna,relevance3}. However, online LLM cross-encoders are often too expensive to run at scale \citep{related-pinterest,relevance-distill,relevance-distill2,rel-distill-reason}, so industrial deployments remain limited to distillation or offline feature generation \citep{related-pinterest,related-ebay,related-walmart,relevance-tencent,chen2024hllm}.

We propose a multi-stage framework that preserves cross-encoder interactions while predicting multiple utilities in one SLM (Figure~\ref{fig:slm_ranking_overall}):
\begin{enumerate}
\setlength{\itemsep}{0pt}
    \item train task-specialized teacher models;
    \item distill from multiple teachers into a unified student (multi-teacher distillation or MTD);
    \item apply targeted optimizations (loss masking, LLM feature engineering, MixLM~\cite{li2025mixlmhighthroughputeffectivellm}).
\end{enumerate}

\paragraph{Prompt Structure}
We use structured prompts for pointwise ranking: \text{system prefix}+\text{context}+\text{document}+\text{suffix}.
Prefix/suffix define the chat template and task instructions; the context contains the query and searcher-side features (searcher profile, interactions). For a fixed query, context is shared across all candidates.

\begin{figure}
    \centering
    \includegraphics[width=1.0\linewidth]{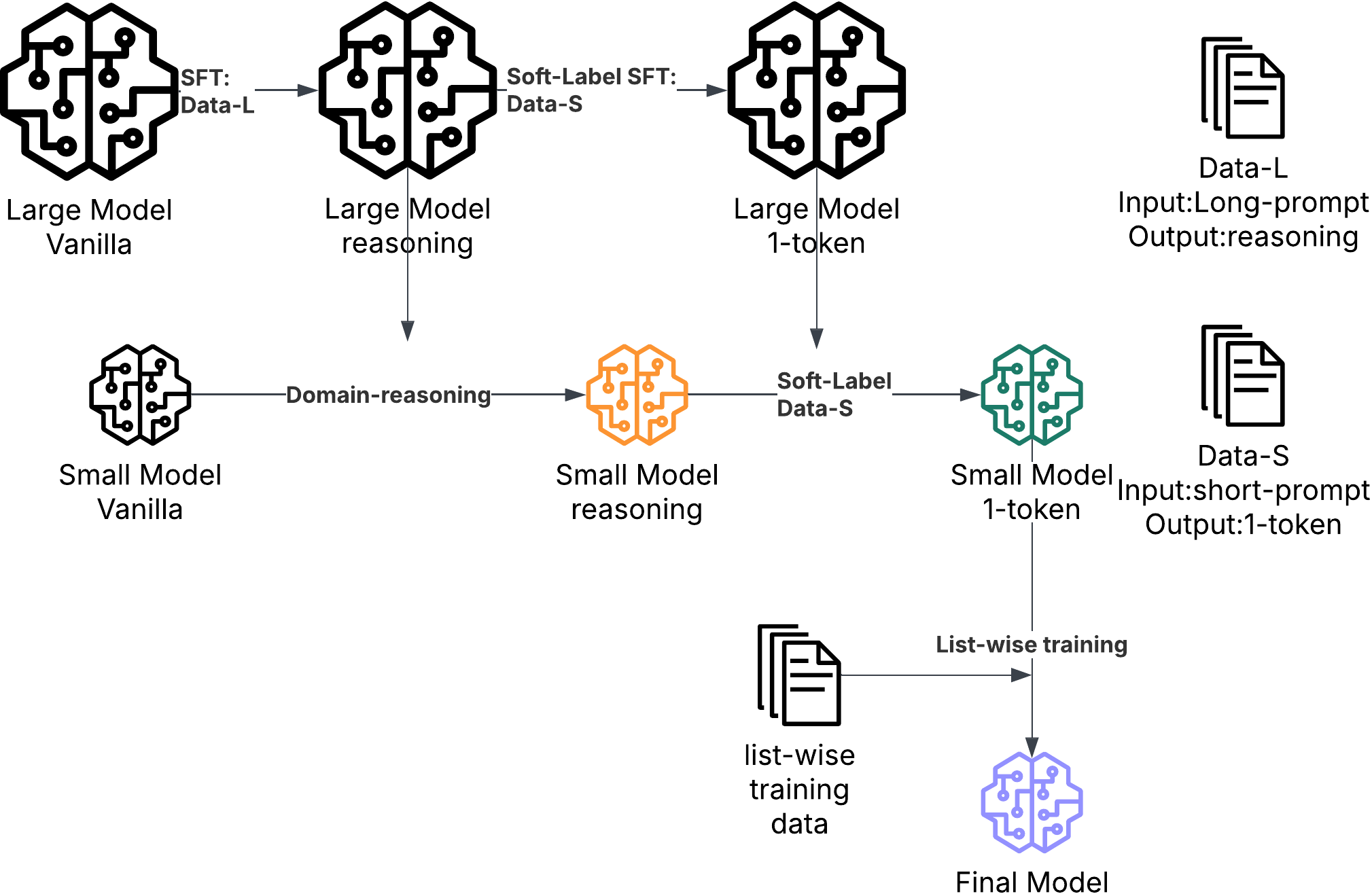}
    \captionsetup{font=small}
    \caption{Relevance Training of SLM Ranker.}
    \vspace{-2mm}
    \label{fig:system_diagram_relevance}
\end{figure}

\subsubsection{\textbf{Relevance Training}}\label{subsubsec:relevance_teacher}
Relevance training/evaluation relies on an 8B ``oracle'' relevance model serving as a policy-aligned judge. Given a query and a candidate document representation (structured attributes plus selected text), it produces an ordinal relevance grade used to supervise retrieval and ranking.

\paragraph{Relevance-only SLMs (Tables~\ref{tab:sjs-ndcg-ablation} and~\ref{tab:sps-ndcg-ablation}).}
We first train a small language model (SLM) whose sole objective is to predict the oracle’s relevance signal (i.e., a \emph{relevance-only} model). Table~\ref{tab:sjs-ndcg-ablation} summarizes this training recipe; we highlight four ingredients that materially impact quality and calibration:
\begin{itemize}
    \item \textbf{Domain Reasoning Distillation:}
    Initialize from an intermediate base model and distill from the oracle via distribution matching (e.g., forward/reverse KL), producing ordinal grades with reasoning text to transfer domain-specific decision rules.
    \item \textbf{Soft-label Fine-Tuning:}
    Map oracle grades to fixed targets in $[0,1]$ and train binary \texttt{Yes}/\texttt{No} probabilities (no rationales), preserving uncertainty near the decision boundary. Soft-label SFT achieves $0.8420$ NDCG@10 vs.\ $0.7583$ with ordinal labels ($+11.0\%$) for Job search. Note that the mapping function makes a difference too. For example, moving to a sigmoid mapping from a linear mapping drove a $+0.28\%$ NDCG lift ($0.8608 \rightarrow 0.8632$).
    \item \textbf{Ranking Loss:}
    Fine-tune on list-wise data: for each query, include top-$k$ EBR documents plus $k$ random documents; form ordered pairs (Equation~\eqref{eq:pairwise_loss}) using oracle scores and optimize a pairwise loss.
    \item \textbf{Chat-template Inference Interface:}
    Use a chat-template system message specifying the binary relevance question and a user message combining the query with a compact candidate representation (title, company, location, seniority, curated text). At inference, we extract first-token \texttt{Yes}/\texttt{No} logits and convert them into probabilities to match the SLM serving interface.
\end{itemize}

\paragraph{Teacher-student upgrade and scaling.}
After establishing the above SLM baseline, we made two key changes.

\textbf{(1) Introducing a relevance teacher.}\label{sec:relevance_teacher}
We train an intermediate \emph{relevance teacher} using the same recipe above, but with a larger model. The teacher is trained directly on the oracle signal (including the ordinal-to-$[0,1]$ mapping for soft-labeling), and then used for high-throughput inference. We subsequently train the student 0.6B SLM \emph{on the teacher’s calibrated \texttt{Yes}/\texttt{No} soft labels}, rather than training the 0.6B model directly against ordinal-mapped soft targets from the oracle. This decouples expensive oracle supervision from student training, improves throughput, and yields a better-calibrated student under the binary \texttt{Yes}/\texttt{No} interface.

\textbf{(2) Data scaling to saturation.}
We scale labeled data volume by $10\times$ for People Search and $40\times$ for Job Search (200k $\rightarrow$ 8M query--document pairs). Beyond this point we observe no additional relevance gains from further increasing dataset size, suggesting performance has saturated for this modeling/setup regime; further improvements likely require changes to the model or supervision strategy.

\begin{table}[t]
\centering
\small
\captionsetup{font=small}
\caption{Ablation study for the ``relevance-only'' student SLM in Job Search.}
\vspace{-4mm}
\begin{tabular}{lcc}
\hline
Techniques & NDCG@10 & $\Delta$ vs Baseline \\
\hline
Open Source SLM & -- & -- \\
+ Ordinal labels SFT  & 0.7583 & \textit{baseline} \\
+ Soft-label SFT (linear mapping) & 0.8420 & +11.04\% \\
+ Domain reasoning & 0.8500 & +12.09\% \\
+ Schedule-free \& per-layer LR tuning & 0.8608 & +13.52\% \\
+ Soft-label SFT (sigmoid mapping) & 0.8632 & +13.83\% \\
+ Chat template & 0.8718 & +14.97\% \\
+ Ranking loss (Formula \eqref{eq:pairwise_loss}) & 0.8772 & +15.68\% \\
+ Base model upgrade ($0.5B \rightarrow 0.6B$) & 0.8910 & +17.50\% \\
+ Distillation from Relevance Teacher & 0.8950 & +18.03\% \\
+ 40$\times$ data volume & \textbf{0.9432} & +24.48\% \\
+ Summarized job descr. + Model pruning & \textbf{0.9218} & +21.56\% \\
\hline
\end{tabular}
\label{tab:sjs-ndcg-ablation}
\end{table}


\begin{table}[t]
\centering
\small
\captionsetup{font=small}
\caption{Ablation study for the ``relevance-only'' student SLM in People Search.}
\vspace{-4mm}
\setlength{\tabcolsep}{3pt}
\begin{tabular}{lcc}
\hline
Techniques & NDCG@5 & $\Delta$ vs Baseline \\
\hline
Open Source SLM & 0.8123 & \textit{baseline} \\
+ Domain reasoning & 0.8290 & +2.05\% \\
+ Soft-label SFT & 0.8410 & +3.53\% \\
+ Chat template & 0.8483 & +4.43\% \\
+ Ranking loss & 0.8890 & +9.44\% \\
+ Distillation from Relevance Teacher & \textbf{0.8933} & \textbf{+9.97\%} \\
\hline
\end{tabular}
\label{tab:sps-ndcg-ablation}
\end{table}

\subsubsection{\textbf{Multi-task Engagement Teacher}}\label{subsec:engagement teacher}
Similar to the relevance teacher (\S\ref{sec:relevance_teacher}), we train an engagement teacher specialized for multiple engagement objectives (e.g., click/apply/dismiss in Job search; long-dwell/connect/follow/message in People Search). Starting from an open-source 1.7B model, we use supervised fine-tuning on action logs. Bundling data engineering, feature engineering and hyperparameter tuning, we improved Job Click AUROC by 4.4\% in comparison to the baseline~\cite{LiRank_paper} (Table~\ref{tab:click-auroc-comparison_features}). 
During engagement modeling, we observe that data and feature engineering, as well as
hyper-parameter tuning, substantially improve performance. Data engineering---including
training on fresher data and stratified sampling to balance labels---yields a 1.48\% Job Click
AUROC lift over the LLM baseline. Feature engineering provides larger gains: incorporating
member text features (e.g., headlines, past positions, locations) adds +4.04\% Job Click AUROC,
while activity history features (text sequence of interacted documents, e.g. last 10 job titles member interacted with) contribute an additional
+2.03\%. The member summary feature (\S\ref{sec:seeker_summary}) further improves Job Click AUROC
by 0.7\%. Hyper-parameter tuning yields further gains: halving the peak learning rate
($1\mathrm{e}{-5}\!\rightarrow\!5\mathrm{e}{-6}$) improves Job Click AUROC by +1.73\%, and doubling
the batch size (4$\rightarrow$8) adds +1.23\%. Increasing training data by 28\% provides an
additional +0.39\% lift. 

\begin{table}[t]
\centering
\small
\captionsetup{font=small}
\caption{Ablation study for the engagement teacher model on Click prediction in Job Search. Relative deltas (\%) w.r.t. LiRank\cite{LiRank_paper} and the LLM baseline.}
\vspace{-4mm}
\begin{tabular}{lccc}
\hline
Method & \makecell{Click\\AUROC} & \makecell{$\Delta$ vs \\ LiRank} & \makecell{$\Delta$ vs \\ LLM Baseline} \\
\hline
\makecell{LiRank\cite{LiRank_paper}: DCNv2+TransAct+GNN} & 0.6487 & \textit{baseline} & -- \\
LLM baseline: query-job only & 0.6069 & -6.45\% & \textit{baseline} \\
+ Fresh data \& balanced actions  & 0.6158 & -5.07\% & +1.48\% \\
+ Member profile text & 0.6403 & -1.29\% & +5.52\% \\
+ Activity history text & 0.6527 & +0.61\% & +7.55\% \\
+ Learning rate: 1e-5 -> 5e-6 & 0.6632 & +2.23\% & +9.28\% \\
+ Doubled batch size (4->8) & 0.6706 & +3.38\% & +10.51\% \\
+ 28\% more training data & 0.6730 & +3.75\% & +10.90\% \\
+ Member summary (\S\ref{sec:seeker_summary}) & \textbf{0.6772} & \textbf{+4.40\%} & \textbf{+11.60\%} \\
\hline
\end{tabular}
\label{tab:click-auroc-comparison_features}
\end{table}

Unlike the relevance teacher (next-token \texttt{Yes}/\texttt{No} prediction), the engagement teacher predicts user actions. Given multiple types of user actions, the engagement teacher optimizes for multiple objectives, thus it requires explicit trade-offs. Table~\ref{tab:multitask-auroc-compact_loss_weight} summarizes loss-weighting effects among 5 tasks in Job Search.

\begin{table}[t]
\centering
\small
\setlength{\tabcolsep}{3pt}
\vspace{-4mm}
\captionsetup{font=small}
\caption{Multi-task AUROC under different loss-weighting strategies in Job Search. Relative (\%) deltas after | w.r.t. the LiRank\cite{LiRank_paper} baseline.}
\vspace{-4mm}
\begin{tabular}{lcll}
\hline
Task & Baseline~\cite{LiRank_paper} & Equal weights & Custom weights\footnotemark \\
\hline
Badfit   & 0.778 & 0.928 |+19.28\% & 0.914 |+17.48\% \\
Apply    & 0.792 & 0.796 |+0.50\%  & 0.804 |+1.56\% \\
Click    & 0.648 & 0.672 |+3.61\%  & 0.677 |+4.40\% \\
Dismiss  & 0.956 & 0.948 |-0.81\%  & 0.943 |-1.32\% \\
Shortlist& 0.801 & 0.898 |+12.11\% & 0.898 |+12.11\% \\
\hline
\end{tabular}
\label{tab:multitask-auroc-compact_loss_weight}
\end{table}

\footnotetext{Custom weights:
\{'click': 0.4, 'apply': 0.4, 'badfit': 0.05, 'shortlist': 0.05, 'dismiss': 0.1\}.}

\subsubsection{\textbf{Engagement and Relevance Joint Optimization}}
\paragraph{Multi-teacher Distillation (MTD)}\label{subsec:mtd}
Training a student model to jointly optimize six objectives—yes/no relevance token prediction and five engagement classification tasks—from two larger teacher models presents two main challenges. First, the smaller student model has limited capacity to retain all knowledge with minimal performance loss. Second, the student must balance potentially conflicting signals from multiple teachers.

During training, each batch is forwarded through teachers to obtain task-specific logits, and the student distills from teacher distributions via weighted Kullback–Leibler (KL) divergence losses. This process preserves teacher uncertainty that hard binarized labels can not offer. The weighted summation of KL losses also allows emphasis on challenging or application-specific tasks.

Single-stage distillation from an open-source checkpoint is inefficient for achieving near-teacher performance across a full spectrum of tasks. Thus, we warm-start from a relevance-specialized SLM, which largely retains the performance of the relevance teacher (as described in \S\ref{sec:relevance_teacher}). Subsequently, MTD is applied to distill the student model from all teachers. This initialization strategy allows MTD to focus primarily on learning engagement objectives, while assigning a small loss weight to prevent drift in relevance performance. Compared to initializing from an open-source model, warm-start initialization gives an improvement of +0.68\% NDCG@10 and +1.98\% in Click AUROC in Job Search (Table~\ref{tab:mtd-init-experiments}).

\begin{table}[t]
\centering
\small
\captionsetup{font=small}
\setlength{\tabcolsep}{3pt}
\caption{Ablation of student model initialization in Multi-Teacher Distillation (MTD). Relevance teacher NDCG@10 = 0.9484. Engagement teacher Click AUROC = 0.6772. $\Delta$Rel and $\Delta$Eng denote relative gap to the relevance and engagement teachers, respectively.}
\vspace{-4mm}
\begin{tabular}{lcccc}
\hline
Student Initialization & NDCG@10 & $\Delta$Rel & Click AUROC & $\Delta$Eng \\
\hline
\makecell[l]{Initialized from\\an open-source model} & 0.9088 & -4.18\% & 0.6574 & -2.92\% \\
\makecell[l]{Warm-start from\\a relevance-specialized SLM} & \textbf{0.9150} & \textbf{-3.52\%} & \textbf{0.6704} & \textbf{-1.00\%} \\
\hline
\end{tabular}
\label{tab:mtd-init-experiments}
\end{table}

Joint-optimization of relevance and engagement via MTD also benefits from data and feature engineering.
In People Search, using stratified sampling and increasing data volume result in a +6\% lift in Follow AUROC, while incorporating extra document, searcher and network features (e.g. past impressions, clicks, network size, distance, number of common connections) give a +17.6\% increase in Connect AUROC (Table~\ref{tab:multitask-auroc-sps}).

\paragraph{Loss Masking}
In People Search, rare actions like Follow/Message are far sparser than Clicks/Long-dwells/Connects. Joint pointwise training can treat positives for other actions as negatives, collapsing rare-action probabilities and complicating score composition.
We therefore train each action head with positives from documents that received the action and negatives from documents shown alongside positives for the same query.
Even if this technique is expected to negatively impact AUROC, it does not affect ranking metrics and results in predicted scores that are $5\times$ greater on average for rare events in comparison to the baseline.
More details in Appendix~\ref{sec:Label_masking_appendix}.

\begin{table}[t]
\centering
\small
\setlength{\tabcolsep}{2pt}
\renewcommand{\arraystretch}{1.05}
\captionsetup{font=small}
\caption{Ablation study for the multi-task engagement SLM model in People Search (relevance + engagement).
Relative (\%) deltas after | w.r.t.\ the engagement SLM query-document baseline.}
\vspace{-4mm}
\setlength{\tabcolsep}{2pt}
\begin{tabular}{lllll}
\hline
\makecell[l]{Method} & Long-dwell & Connect & Follow & Message \\
& AUROC & AUROC & AUROC & AUROC \\
\hline
\makecell[l]{Relevance SLM:\\query-doc only} & .605 & .558 & .494 & .513 \\
\makecell[l]{Engagement SLM:\\query-doc only} & .676 & .727 & .815 & .833 \\
\makecell[l]{+ Balanced actions}
& .673|$-0.5\%$ & .723|$-0.6\%$ & .823|+1.1\% & .819|$-1.7\%$ \\
\makecell[l]{+ Balanced queries}
& .671|$-0.8\%$ & .727|+0.0\% & .831|+2.0\% & .827|$-0.8\%$ \\
\makecell[l]{+ 10$\times$ training data}
& .701|+3.8\% & .767|+5.5\% & .865|+6.0\% & .861|+3.4\% \\
\makecell[l]{+ Document\\numeric features}
& .718|+6.2\% & .774|+6.5\% & .884|+8.2\% & .874|+4.9\% \\
\makecell[l]{+ Searcher features}
& .771|+13.6\% & .831|+13.8\% & .908|+10.9\% & .903|+8.2\% \\
\makecell[l]{+ Network features} & \textbf{.797|+16.9\%} & \textbf{.908|+23.1\%} & \textbf{.923|+12.6\%} & \textbf{.933|+11.5\%} \\
\hline
\end{tabular}
\label{tab:multitask-auroc-sps}
\end{table}

\subsubsection{\textbf{Calibration}}
Deep ranking scores are often miscalibrated due to discriminative losses and training-time sampling / downsampling \cite{pmlr_v70_guo17a}. We apply a post-processing calibration layer: a PyTorch isotonic-regression-style model \cite{LiRank_paper} augmented with feature embeddings, which ingests raw SLM scores plus context and outputs calibrated probabilities and derived signals. The design is multi-head and modular: each head is trained independently with its own features and calibration labels (primary, proxy, business-aligned) and composed into a single serving artifact. It also supports position-conditioned calibration via loss masking, learning rank-specific mappings for position-aware ranking/auction logic.

We deploy two calibration modes in one artifact: a global probability and a position-conditional vector $\{\hat{p}^{(r)}\}_{r=1}^{25}$, where $\hat{p}^{(r)}$ is the outcome likelihood if placed at rank $r$. This is implemented as additional loss-masked outputs in the same forward pass (no extra online stage). Position conditioning improves Click AUROC from 0.6704 to 0.7095 on top of multi-teacher distilled SLM in Job Search, consistent with exposure bias and motivating position-aware allocation (e.g., VCG \cite{vcg_paper}). Calibration also improves probability fidelity: O/E ratios move toward 1.0 from a broad pre-calibration spread.

\subsubsection{\textbf{Feature Engineering with SLMs}}

\paragraph{Summarization of Member Profile and History with Raw Text}\label{sec:seeker_summary}
Because member context is heterogeneous, we build a unified member summary from profiles, professional content, search activity, and historical queries. Naive summarization degrades performance, so we train a summarizer with RL similar to \cite{lin2025recr1bridginggenerativelarge}, see more details in \cite{summarization_user, behdin2025scaling}. An open-source 1.7B model is the actor; the engagement teacher (\S\ref{subsec:engagement teacher}) is the reward model. Given a time-ordered sequence
\[
\langle d_1, a_1 \rangle, \ldots, \langle d_n, a_n \rangle,
\]
of documents ($d_i$) and actions ($a_j$)
we reserve $\langle d_n, a_n\rangle$ for reward computation. The summary should suffice for the teacher to predict $a_n$ for $d_n$. For a summary $s$, we combine prediction reward with length and quality terms:
\begin{equation}
R(s) =
\mathbb{I}[\hat{a}_n = a_n]\Big(
1
- \lambda_{\text{len}} \cdot \ell(s)
+ \lambda_{\text{qual}} \cdot q(s)
\Big),
\end{equation}
where $\ell(s)$ is a normalized length penalty and $q(s)$ is a factuality/saliency score from a 32B model; if $\hat{a}_n \neq a_n$, $R(s)=0$. Training uses GRPO \cite{shao2024deepseekmath} with clip-higher \cite{yu2025dapo} and no standard deviation in advantage estimation \cite{liu2025understanding}.
Using the searcher profile summary on top of the query for Job search improves engagement-head AUROCs Apply and Shortlist by $\sim$1\% each. For People Search, we use summarized profiles to represent both searchers and documents.

\paragraph{Numerical Features}
Beyond history, engagement depends on additional signals (e.g., common connections, network distance, historical CTR). Table~\ref{table:numerical-features} shows that prompts with descriptive feature names, boolean True/False encoding, and explicit CTR (with clicks/impressions) improve AUROC, while truncation to two decimals reduces tokens without harming performance.

\begin{table}[h]
\centering
\small
\captionsetup{font=small}
\caption{Experiments on ways of incorporating numerical feature in prompt; $\Delta$ denotes incremental gains over the previous lines.}
\vspace{-4mm}
\setlength{\tabcolsep}{3pt}
\begin{tabular}{lcc}
\hline
Variant & $\Delta$ in AUC \\
\hline
Short feature identifiers & \textit{baseline} \\
+Descriptive feature identifiers & +5.8\% \\
+Binary feature values as True/False & +1.7\% \\
+CTR feature & +5.1\% \\
+Truncation to first 2 decimal places & 0.0\% \\
\hline
\end{tabular}
\label{table:numerical-features}
\end{table}





\subsection{Semantic Search Retrieval}\label{subsection:semantic-search-retrieval}
Retrieval is the first stage of Semantic Search, producing a high-recall candidate set for
downstream ranking. Our design builds on LLM-based embedding models, now the dominant approach
for large-scale semantic retrieval. We use a contrastive LLM bi-encoder to map queries and
documents into a shared representation space, enabling efficient nearest-neighbor search
while capturing semantic similarity beyond lexical overlap. Prior work shows that
decoder-only LLMs can be turned into strong embedders via contrastive objectives and
bidirectional attention \cite{llm2vec}, and that limited embedding-specific fine-tuning
suffices to obtain universal embeddings \cite{universal_embedders}. Retrieval quality also depends critically on the training signal and data construction. Effective
hard-negative mining improves discrimination at top ranks \cite{nvretriever}, while
multilingual and task-adaptation techniques (e.g., task LoRA) provide robustness across
heterogeneous query distributions \cite{jina_embv3}. We adopt these principles to ensure high
recall and generalization.

Given a query $q$ and a corpus of up to 1.3B documents, the retrieval layer selects 
the best top-$K$ candidates ($K{=}1000$). Our system combines fine-tuned embedding models with
in-GPU \emph{retrieval-as-ranking} (RAR) distance models \cite{borisyuk2024linrmodelbasedneural}
to score the full corpus while jointly optimizing relevance and engagement.


\subsubsection{\textbf{Data Engineering}}
Training and evaluation data are derived from LinkedIn search traffic annotated by the relevance 8B LLM ``oracle'' (\S\ref{subsubsec:relevance_teacher}), yielding $\langle q,d\rangle$ pairs with graded relevance labels $\{1,2,3,4\}$. We improve data
quality via high-confidence filtering and de-duplication, and adopt query-centric hard-negative
sampling: for each query, we sample 1–2 positives ($label>2$) and 2–3 hard negatives ($label \leq 2$) drawn from
top-ranked production candidates labeled non-relevant by the LLM judge. This preserves query
diversity and emphasizes real-world top-rank failure modes.

Training data is constructed via policy-bucketed sampling and
hard-negative mining. Policy sampling consists of bucketing queries into semantic categories using
an LLM-based tagger and then re-sizing bucket $i$ ($B_i$) based on product needs ($P_i$),
as well as the bucket's offline quality gap between the baseline and treatment retrievers ($G_i$), measured by:
\begin{equation}\label{eq:quality_upsampling}
G_i = \frac{\mathrm{Precision@10}_{\text{baseline}, i}}{\mathrm{Precision@10}_{\text{treatment}, i}}, \qquad
B_i = P_i \cdot G_i.
\end{equation}
Contrastive tuples are formed via rank-aware hard-negative mining, combining
positives, hard negatives, easy negatives, and global negatives to emphasize informative decision boundaries.

\subsubsection{\textbf{Modeling}}
Our retrieval model is an LLM-based bi-encoder trained with contrastive learning.  
Given query $q$ and document $d$, we denote their embeddings as $e_q$ and $e_d$, and score them via cosine similarity $\langle e_q, e_d \rangle = \cos(e_q, e_d)$.  

Training combines a global InfoNCE objective ($\mathcal{L}_{\text{InfoNCE}}$) with a pairwise margin loss ($\mathcal{L}_{\text{pair}}$) to sharpen local decision boundaries:
\begin{equation}\label{eq:info_loss}
\mathcal{L}_{\text{InfoNCE}} = -\log 
\frac{\exp\left(\langle e_q,e_{d^{+}} \rangle / \tau\right)}
{\exp\left(\langle e_q,e_{d^{+}} \rangle / \tau\right) + 
\sum_{d^{-} \in \mathcal{B}^{-}} \exp\left(\langle e_q,e_{d^{-}} \rangle / \tau\right)},
\end{equation}
where $\tau>0$ is a temperature hyperparameter that controls the sharpness of the contrastive distribution, and $\mathcal{B}^{-}$ includes both in-batch negatives and explicitly mined hard negatives,
\begin{equation}\label{eq:pairwise_loss}
\mathcal{L}_{\text{pair}} = \sum_{d^{-} \in \mathcal{D}^{-}} 
\max\left(0,\, m - \langle e_q,e_{d^{+}} \rangle + \langle e_q,e_{d^{-}} \rangle \right),
\end{equation}
where $\mathcal{D}^{-}$ denotes a curated set of hard negatives for each query (e.g., retrieved but non-relevant documents), and $m>0$ is a margin that enforces a minimum separation between positives and hard negatives.

The final objective
$\lambda\ \mathcal{L}_{\text{InfoNCE}} + (1-\lambda)\ \mathcal{L}_{\text{pair}}$
preserves global semantic structure while resolving subtle constraint violations at top ranks.

\subsubsection{\textbf{GPU RAR Model Training}}
While the bi-encoder captures semantic relevance distilled from a relevance-focused teacher,
retrieval must also reflect personalization and engagement preferences. We therefore replace
pure cosine similarity scoring with GPU retrieval-as-ranking (GPU RAR) scoring
\cite{borisyuk2024linrmodelbasedneural}:
\begin{equation}
S(q,d) = w_0 \langle e_q,e_d\rangle + \sum_{i=1}^{n} w_i f_i(q,d),
\end{equation}
where $f_i$ are personalization and engagement features (e.g., network proximity, profile popularity). 
GPU RAR is trained with a weighted multi-task objective:
\begin{equation}
\mathcal{L_{\mathrm{RAR}}} = \lambda\,\mathcal{L}_{\mathrm{BCE}}(S, L_R)
+ (1-\lambda)\,\mathcal{L}_{\mathrm{BCE}}(S, L_E),
\end{equation}
where $L_R$ and $L_E$ denote relevance and engagement labels.

\subsubsection{\textbf{GPU Retrieval Serving}}
We deploy optimized query and document embedding towers on a GPU-accelerated exhaustive retrieval stack with attribute-based pre-filtering~\cite{yuchin10.1145/3705328.3748116, borisyuk2024linrmodelbasedneural}. To meet throughput and robustness requirements, we reduce embedding dimensionality, support two-slot multi-model deployment for controlled A/B ramps, enforce vRAM guardrails during internationalization and backfills, and validate new models via counterfactual server-client testing under production-like conditions.

\subsubsection{\textbf{Experimental Results}}
Models are evaluated using embedding-level metrics (Recall@K with breakdowns by query frequency, category, and language) and system-level counterfactual retrieval tests focused on NDCG@K. As shown in Table~\ref{table:ebr_hardneg_results}, InfoNCE fine-tuning substantially improves recall and ranking quality over a strong 8B baseline in Job search, while query-centric hard negatives yield further gains. Full-parameter fine-tuning performs best. A 4B model trained with the same objective closely matches the 8B FPFT variant, indicating effective transfer to smaller models and a favorable accuracy-efficiency trade-off for production.

In People Search (Table~\ref{table:retrieval_p10_r10_ndcg10}), the use of a chat template, hard-negative mining, and quality-based upsampling substantially improves all metrics over the baseline. GPU RAR further balances relevance and engagement, improving Click AUC by +1.7\% (0.595$\rightarrow$0.603) within the retrieval distance model and relevance NDCG@10 by +1.28\% (0.78$\rightarrow$0.79). By explicitly modeling social network proximity as features in the model, GPU RAR replaces multiple production retrieval paths that were previously split across separate calls for different network distances with a single retrieval call, reducing system complexity and infrastructure cost.


\begin{table}[tb]
\captionsetup{font=small}
\caption{Retrieval model metrics (precision, recall, and NDCG at 50) for Job search.}
\vspace{-4mm}
\setlength{\tabcolsep}{3pt}
\centering
\small
\begin{tabular}{lccc}
\hline
Model Variant & P@50 & R@50 & NDCG@50 \\
\hline
Baseline-8B & 0.414 & 0.774 & 0.735 \\
+ Chat template & 0.446 & 0.830 & 0.788 \\
+ InfoNCE (LoRA) & 0.471 & 0.874 & 0.829 \\
+ InfoNCE + HardNeg (LoRA) & 0.497 & 0.887 & 0.833 \\
+ InfoNCE + HardNeg (FPFT) & \textbf{0.505} & \textbf{0.899} & \textbf{0.842} \\
4B + InfoNCE + HardNeg (FPFT) & 0.501 & 0.889 & 0.834 \\
\hline
\end{tabular}
\label{table:ebr_hardneg_results}
\vspace{-3mm}
\end{table}

\begin{table}[tb]
\captionsetup{font=small}
\caption{Retrieval model metrics (precision, recall and NDCG at 10) for People Search.}
\vspace{-4mm}
\setlength{\tabcolsep}{3pt}
\centering
\small
\begin{tabular}{lccc}
\hline
Model Variant & P@10 & R@10 & NDCG@10 \\
\hline
Baseline-4B & 0.33 & 0.70 & 0.71 \\
+ Chat template & 0.36 & 0.74 & 0.74 \\
+ HardNeg Mining & 0.41 & 0.75 & 0.76 \\
+ Quality based upsampling Eq\eqref{eq:quality_upsampling} & 0.46 & 0.78 & 0.78 \\
+ GPU RAR model & 0.47 & 0.79 & 0.79 \\
\hline
\end{tabular}
\label{table:retrieval_p10_r10_ndcg10}
\vspace{-3mm}
\end{table}

\subsection{Modeling for SLM Inference Optimizations}

\begin{table}[t]
\captionsetup{font=small}
  \caption{Efficiency comparison at a fixed 500ms latency budget, reporting NDCG@10 and throughput items/s on a single H100.}
  \vspace{-4mm}
  \setlength{\tabcolsep}{2pt}
  \small
  \label{tab:sota2}
  \begin{tabular}{lrrr}
    \hline
    Model & NDCG@10 & QPS (items/s/GPU) & Latency (ms) \\
    \hline
    Full Text  & 0.9432 & 290 & $<500$ \\
    Summarized + Pruned & 0.9218 & 2{,}200 & $<500$ \\
    MixLM~\citep{li2025mixlmhighthroughputeffectivellm} & 0.9239 & 22{,}000 & $<500$ \\
    \hline
  \end{tabular}
\end{table}

Serving cross-encoder SLM rankers at scale is challenging due to large model size and long input contexts. To satisfy strict latency and throughput constraints, we apply complementary modeling optimizations that substantially reduce inference cost while preserving ranking quality (Table~\ref{tab:sota2}).

\subsubsection{\textbf{Context Summarization}}
Inference cost scales linearly with input length, so we compress long text fields especially for documents (e.g., job descriptions in Job Search, member profiles in People Search) via offline summarization. Summaries are generated using a larger 1.7B LLM and stored for online serving, reducing document text length by an order of magnitude with negligible quality loss. In Job Search, this yields a 4$\times$ throughput improvement; in People Search, it reduces 95th-percentile prompt length from $\sim$1{,}500 to $\sim$500 tokens without measurable degradation. The summarization model is trained using reinforcement learning, see more in \citep{behdin2025scaling}. 

\subsubsection{\textbf{Model Pruning}}
We apply structured compression using OSSCAR~\citep{meng2024osscar}, pruning 50\% of hidden neurons in each MLP block and removing the final eight transformer layers. This reduces model size from 600M to 375M parameters, followed by fine-tuning to recover quality. As shown in Table~\ref{table:sps_pruning}, the pruned model matches or slightly exceeds the dense baseline in People Search; similar trends hold in Job Search~\cite{behdin2025scaling}. Combined with context summarization, pruning increases throughput from 290 to 2{,}200 items/s/GPU (7.5$\times$) under the same latency budget.

\begin{table}[tb]
\captionsetup{font=small}
\caption{Ranking quality in People Search for dense and pruned SLMs}
\vspace{-4mm}
\centering
\small
\begin{tabular}{lccc}
\hline
Model & Precision@10 & Recall@10 & NDCG@10 \\
\hline
Baseline (600M) & 0.5369 & 0.8597 & 0.8629 \\
Pruned (375M)   & 0.5434 & 0.8578 & 0.8652 \\
\hline
\end{tabular}
\label{table:sps_pruning}
\vspace{-3mm}
\end{table}

\subsubsection{\textbf{Text-Embedding Mix Interaction}}
We further reduce inference cost using MixLM~\citep{li2025mixlmhighthroughputeffectivellm}, a text-embedding mixed-interaction architecture. A dedicated encoder compresses each item into a small set of learned embedding tokens, cached nearline. At inference, the ranker consumes the query text plus these embedding tokens, preserving cross-encoder interactions while reducing context length by orders of magnitude. The encoder and ranker are trained end-to-end via multi-stage distillation against a full-text teacher. In production, MixLM delivers $\sim$10$\times$ higher throughput than summarized-text SLMs and $\sim$76$\times$ over raw-text SLMs at the same latency budget, enabling full-scale LLM-based ranking.

\subsection{Training Infrastructure Optimizations}

Efficient training of the ranking components in \S\ref{SLM_ranking} relies on coordinated system- and algorithm-level optimizations that accelerate teacher fine-tuning, scale multi-teacher distillation, and reduce the cost of history modeling and large-scale GPU training. To speed up multi-task teacher fine-tuning, we adopt LiGer~\cite{hsu2024liger}, which reduces memory usage and enables 2$\times$ larger batch sizes. We further leverage multi-node training (up to 3.5$\times$ speedup), FSDP2~\cite{pytorch_fsdp2} (additional 20\% gain), and H200 multi-node clusters (up to 30\% further reduction). We evaluated FP8 mixed precision but observed no benefit for models $<8$B parameters due to casting overhead.

\subsubsection{\textbf{Multi-Teacher Distillation Framework}}
To effectively train the student model with multiple teachers, as outlined in \S\ref{subsec:mtd}, we developed a specialized framework built on SGLang. This system is capable of loading and serving teacher models of various sizes at runtime, while handling Tensor-Parallelism, and Data Parallelism configurations. During training, an asynchronous client is responsible for querying the teachers, processing their outputs, and integrating them into the distillation losses. We call this approach Online Multi-teacher Distillation. It scales efficiently across multiple nodes using local teacher replicas, achieving a 3$\times$ speedup in the distillation process while maintaining low latency, which is highly beneficial for rapid iteration. To further mitigate the serving overhead during runtime and avoid redundant computation inherent in the online mode, we also implemented a caching mechanism, which is called Offline Multi-teacher Distillation. In this mode, teacher outputs are precomputed and cached on secondary storage (HDFS and/or NFS) and consumed directly during training. This offline approach yields approximately a 35\% reduction in training time and saves about 25\% in total GPU-hours. Although the data-generation phase introduces a one-time upfront cost, it substantially reduces the overall compute requirements when training student models with shared data and teacher models but different configurations, making it significantly more efficient for repeated experimentation.

\subsubsection{\textbf{Agentic GPU Optimizations}}
We develop a GPU optimization agent that analyzes training code and utilization metrics (e.g., SM\_ACTIVE, SM\_OCCUPANCY) to propose efficiency improvements. The agent identifies relevant workflow code via BFS traversal and RAG-based file filtering, then applies LLM-guided configuration tuning (e.g., gradient checkpointing, FSDP settings). Applied to MixLM training~\cite{li2025mixlmhighthroughputeffectivellm}, this reduces training time by 13\% (256 GPU-hours saved on 64 H100s). 

\subsection{ML Inference Infrastructure Optimizations}
\label{sec:ml_inference}

Large-scale semantic ranking with LLM rankers has inference requirements that differ from generative workloads. In Semantic Job and People Search, each query triggers \emph{prefill-only} scoring for hundreds to thousands of candidates under strict tail-latency constraints. Prompts share a long prefix (system instructions, query, searcher context) and a short item-specific suffix; models output scalar scores from final-token logits. There is no iterative decoding, and throughput must sustain millions of item scores per second. In this regime, cost is dominated by prefill compute, tokenization, and CPU-side orchestration, while generative decode paths become overhead \cite{behdin2025scaling,li2025mixlmhighthroughputeffectivellm}. We deploy an inference stack that makes LLM ranking practical at scale (summarized in Table~\ref{tab:kdd_stacked_ablation_375m}):
(1) scoring-specialized prefill execution (no decode/sampling/per-token logprobs); (2) shared-prefix amortization to remove redundant query-prefix compute; (3) CPU/runtime co-design for tokenization, scheduling, and Python overhead; (4) native mixed-input inference for text-embedding interaction regimes; (5) middle tier optimizations.

\begin{table}[t]
\centering
\small
\setlength{\tabcolsep}{5pt}
\renewcommand{\arraystretch}{1.15}
\captionsetup{font=small}
\caption{Throughput gains from inference optimizations for prefill-only LLM ranking using a 375M pruned decoder-only ranker (50 query + 150 item tokens, batch size 50) on a single H100 GPU under p99 $\leq$500ms. Total speedup over baseline \textbf{2.93$\times$}. }
\vspace{-4mm}
\begin{tabular}{lrr}
\hline
Optimization Stage
& Incremental Gain
& Throughput \\
& (items/s/GPU)
& (items/s/GPU) \\
\hline
Baseline & --  & 750 \\
+ Batch tokenization \& batch send & +150 & 900 \\
+ Scoring-only prefill execution & +400 & 1300 \\
+ Python/runtime optimizations & +300 & 1600 \\
+ In-batch prefix caching (IBPC) & +400 & 2000 \\
+ Piecewise CUDA graph (prefill) & +200 & 2200 \\
\hline
\end{tabular}
\label{tab:kdd_stacked_ablation_375m}
\end{table}

\subsubsection{\textbf{Scoring-Optimized Prefill Execution}}
\label{sec:scoring_prefill}

Decoder-only engines are tuned for autoregressive generation and incur overhead for ranking. We introduce a \emph{scoring-optimized prefill path} \cite{behdin2025scaling} that executes a single forward pass, bypasses decoding and sampling, returns only final-token logits, and releases KV state immediately. We also disable unused per-token log-probability computation, consolidate device-host transfers, and overlap CPU postprocessing with GPU execution.
On a 375M pruned ranker, this improves throughput from 900 to 1300 items/s/GPU (+44\%) under p99 $\leq$500\,ms (scoring-only stage in Table ~\ref{tab:kdd_stacked_ablation_375m}).

\subsubsection{\textbf{Shared-Prefix Amortization}}
\label{sec:prefix_amort}

Ranking prompts share a long query prefix across items; recomputing it per item wastes compute. Let $T_q$ be prefix length, $T_i$ suffix length, and $N_i$ ranking depth. Na\"{i}ve prefill scales as:
\[
F_{\text{att}}^{\text{naive}} \propto N_i (T_q + T_i)^2, \quad
F_{\text{lin}}^{\text{naive}} \propto N_i (T_q + T_i),
\]
while amortized prefill yields:
\[
F_{\text{att}}^{\text{amort}} \propto T_q^2 + N_i (2T_iT_q + T_i^2), \quad
F_{\text{lin}}^{\text{amort}} \propto T_q + N_i T_i.
\]

We employ two shared-prefix amortization strategies. \emph{In-batch prefix caching (IBPC)} \cite{behdin2025scaling} ensures the prefix KV is only computed once and shared across the entire batch. Alternatively, \emph{Multi-Item Scoring} concatenates all items into one prompt and uses an attention mask to prevent inter-item attention. For 50 query tokens, 150 item tokens, and batch size 50, shared-prefix amortization improves throughput from 1600 to 2000 items/s/GPU (+25\%) (Table~\ref{tab:kdd_stacked_ablation_375m}).

\paragraph{CUDA graph execution.}
After removing redundant compute, kernel launch overhead dominates. Piecewise CUDA graph execution captures stable prefill segments while allowing dynamic shapes, increasing throughput from 2000 to 2200 items/s/GPU (+10\%) and yielding a cumulative 2.93$\times$ speedup over baseline.

\subsubsection{\textbf{CPU and Runtime Optimizations}}
\label{sec:cpu_runtime_opts}

After GPU-side improvements, bottlenecks shift to tokenization, serving, and scheduling. Batch tokenization and batch send ensure each request maps to a single prefill pass, improving throughput from 750 to 900 items/s/GPU (+20\%). To bypass Python GIL limits, we use a multi-process gRPC design that parallelizes CPU preprocessing across cores; together with scheduler/runtime tuning, this raises throughput from 1300 to 1600 items/s/GPU (+23\%). We stabilize tail latency by freezing Python heaps post-warmup via \texttt{gc.freeze()}.

\subsubsection{\textbf{Mixed-Input Inference and Embedding-Based Scoring}}
\label{sec:mixed_input}

To support MixLM \cite{li2025mixlmhighthroughputeffectivellm}, we extend the engine to natively handle mixed text-embedding inputs while reusing scoring-only prefill, prefix amortization, and runtime optimizations. Embeddings are sent as binary payloads and injected directly into model inputs, remaining backward-compatible with text-only requests.

When items are represented by a single embedding token, GPU compute becomes negligible relative to the shared prefix and throughput is dominated by CPU orchestration. Table~\ref{tab:embedding_ablation} shows single-process serving saturates at $\sim$10k items/s/GPU, while multi-process serving plus CUDA graphs scales to $\sim$22k items/s/GPU (2.2$\times$).

\begin{table}[t]
\captionsetup{font=small}
\caption{Ablation study of infrastructure optimizations under embedding-based item inputs
for a 0.6B-parameter ranker on NVIDIA H100 GPUs (p99 $\leq$500\,ms).}
\vspace{-4mm}
\label{tab:embedding_ablation}
\centering
\small
\begin{tabular}{p{0.6\columnwidth} r}
\hline
Infrastructure Configuration & Throughput \\
& (items/s/GPU) \\
\hline
Single gRPC servicer + single SGLang worker & $\sim$10\,000 \\
Multi-process serving (6 gRPC + 6 SGLang) & $\sim$19\,500 \\
Multi-process serving + CUDA graph & $\sim$22\,000 \\
\hline
\end{tabular}
\end{table}

\subsubsection{\textbf{Middle Tier Serving Optimizations}}

Because relevance scores for a fixed searcher-query-entity tuple are deterministic for a given model, we cache scored entity IDs in a distributed Couchbase-backed store keyed by searcher identity and query signature. Requests first probe the cache to bypass GPU inference on hits and populate on misses. In production, $>$50\% of scoring requests are served from cache, reducing median/mean latency by $\sim$8--10\% and improving tail latency across online and nearline traffic.
To improve efficiency under variable load, a midtier retry policy enforces latency budgets and smooths burstiness, yielding an additional $\sim$10\% throughput gain by reducing redundant GPU computation. We also use adaptive control, including a PID-based Dynamic Scoring Depth controller that adjusts per-query scoring depth: during peaks, average depth drops from 250 to 130 (48\% less per-query GPU compute), while off-peak periods allow deeper scoring to improve quality. Finally, traffic shaping defers latency-insensitive requests into idle windows, improving GPU scheduling efficiency and increasing effective GPU throughput by $\sim$25\%. Together, these techniques balance quality, scalability, and responsiveness under highly variable production load.




\section{Online Deployment and Future Work}

The deployed Semantic Search stack combines GPU-accelerated embedding-based retrieval with a
high-throughput Small Language Model (SLM) ranker, making LLM-based retrieval and ranking
practical at LinkedIn scale. The SLM ranker integrates multi-staged teacher fine-tuning,
multi-teacher distillation, and text and embedding feature engineering (\S\ref{SLM_ranking}),
and is co-deployed with the ML inference optimizations described in
\S\ref{sec:ml_inference}, enabling high-QPS serving while jointly optimizing relevance
and engagement.

Relative to the prior DLRM-style ranking baseline~\cite{LiRank_paper}, the new system delivers
substantial gains across Semantic Job Search and Semantic People Search. In Job Search, we
observe a +7.73\% improvement in NDCG@10 and a -46.88\% reduction in Poor Match Rate@10
(PMR@10), while People Search achieves over 10\% NDCG@10 improvement. From the time of the launch Semantic Search demonstrated over +1.2\% DAU lift.

Future work will focus on deeper personalization using longer-term user context and further
reducing LLM inference cost through adaptive and selective computation, with the goal of
extending these techniques to large-scale search and recommendation systems.


\bibliographystyle{ACM-Reference-Format}
\bibliography{sample-base}

\section{APPENDIX}

\subsection{Loss Masking for Multi-task Optimization}\label{sec:Label_masking_appendix}

In People Search, engagement actions such as follows and messages are much sparser than clicks or connects.
Optimizing all action heads with independent pointwise losses exacerbates imbalance, since
negatives for a given action include both non-interacted documents as well as documents with other
positive actions. This drives predicted probabilities for rare actions toward zero and
complicates score aggregation.

\begin{figure}
    \centering
    \includegraphics[width=1.0\linewidth]{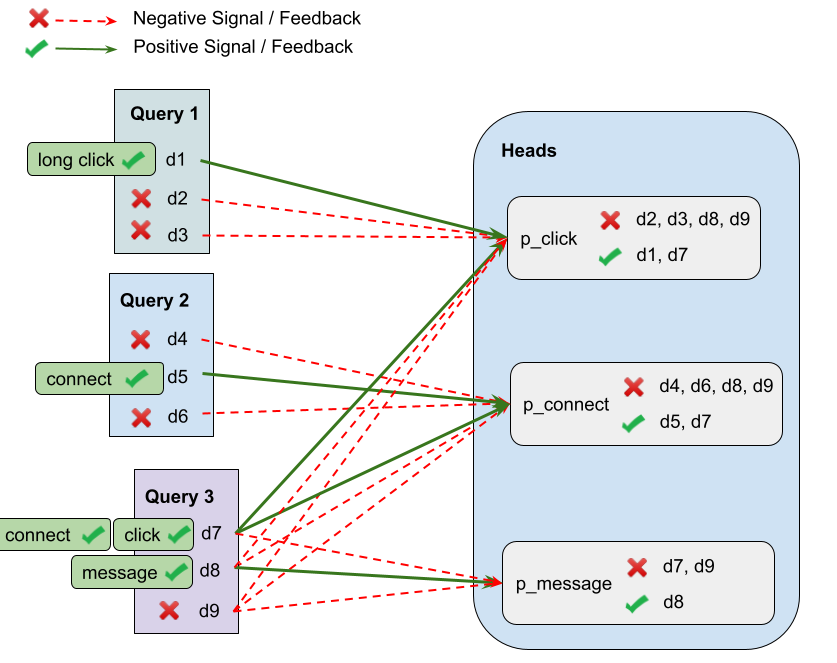}
    \captionsetup{font=small}
    \caption{Negative example selection. Documents are treated as negatives for an action only
    when that action occurs on another document for the same query.}
    \label{fig:loss_masking_diagram_fig}
\end{figure}

To address this, we apply a masked training objective per action head. Positives are documents
that received the target action, while negatives are restricted to documents surfaced for the
same query when at least one document received that action. This generalizes hard-negative
training to a multi-objective setting. In practice, we group data by query, aggregate action binary labels
with a boolean OR, and mask losses accordingly. Figure~\ref{fig:loss_masking_diagram_fig}
illustrates the resulting example selection.

\begin{figure}
    \centering
    \includegraphics[width=1.0\linewidth]{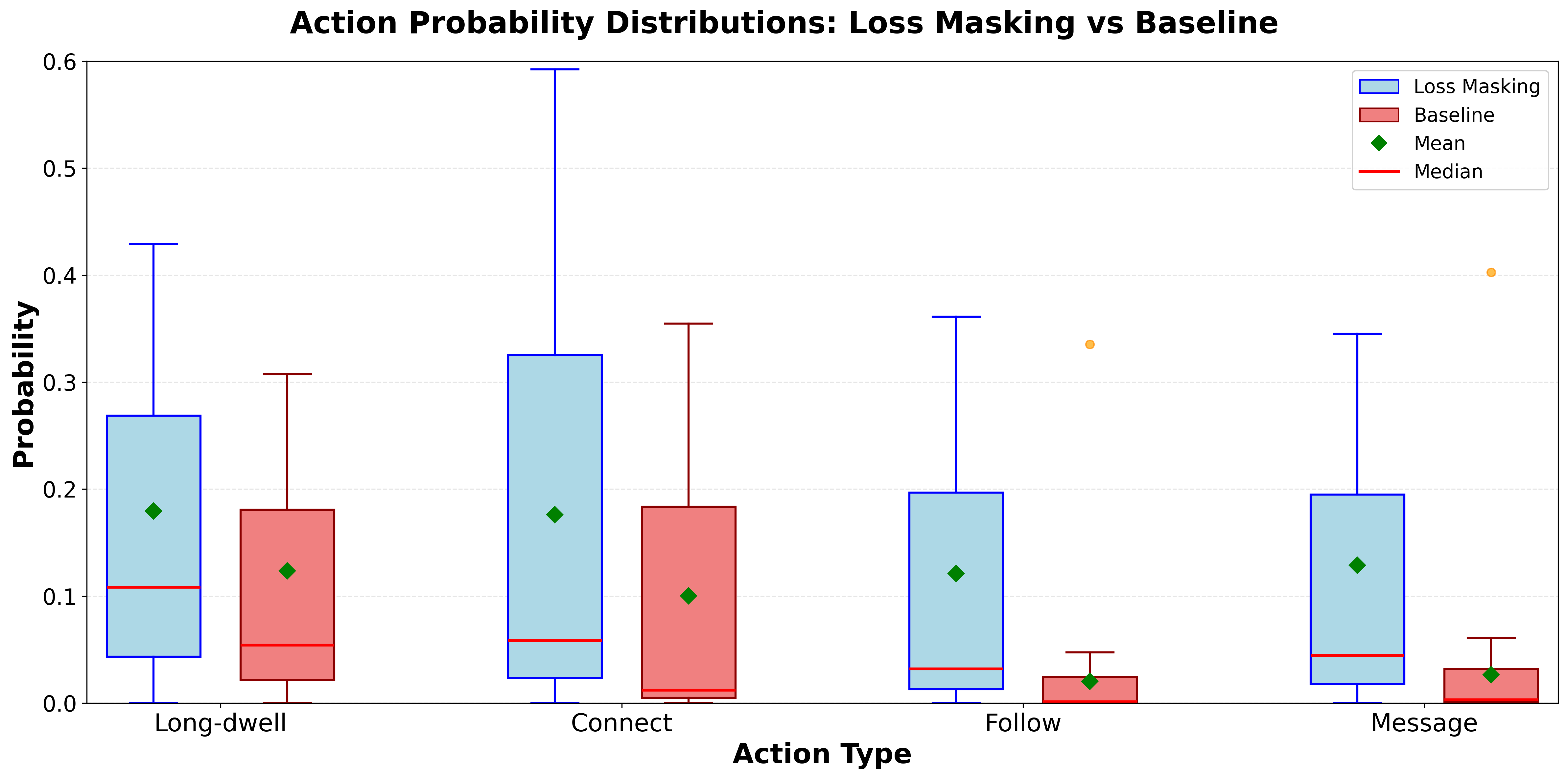}
    \captionsetup{font=small}
    \caption{Distribution of predicted probability scores for the multi-task engagement People Search SLM when trained with loss masking versus an unmasked loss (baseline).}
    \label{fig:loss_masking_dist_diagram_fig}
\end{figure}

When we rank documents with a weighted combination of predicted action scores and compute NDCG against each event (assuming a document is relevant if it received a certain engagement action) we observe non-statistically significant differences between training with and without loss masking. Figure~\ref{fig:loss_masking_dist_diagram_fig} shows the predicted score distribution for each head.
Applying loss masking results in increased absolute predicted scores for heads representing rare events (Follow and Message), while keeping ranking metrics neutral.

\subsection{System Architecture for Semantic Search}\label{sec:system_diag_appendix}

\begin{figure}
    \centering
    \includegraphics[width=1\linewidth]{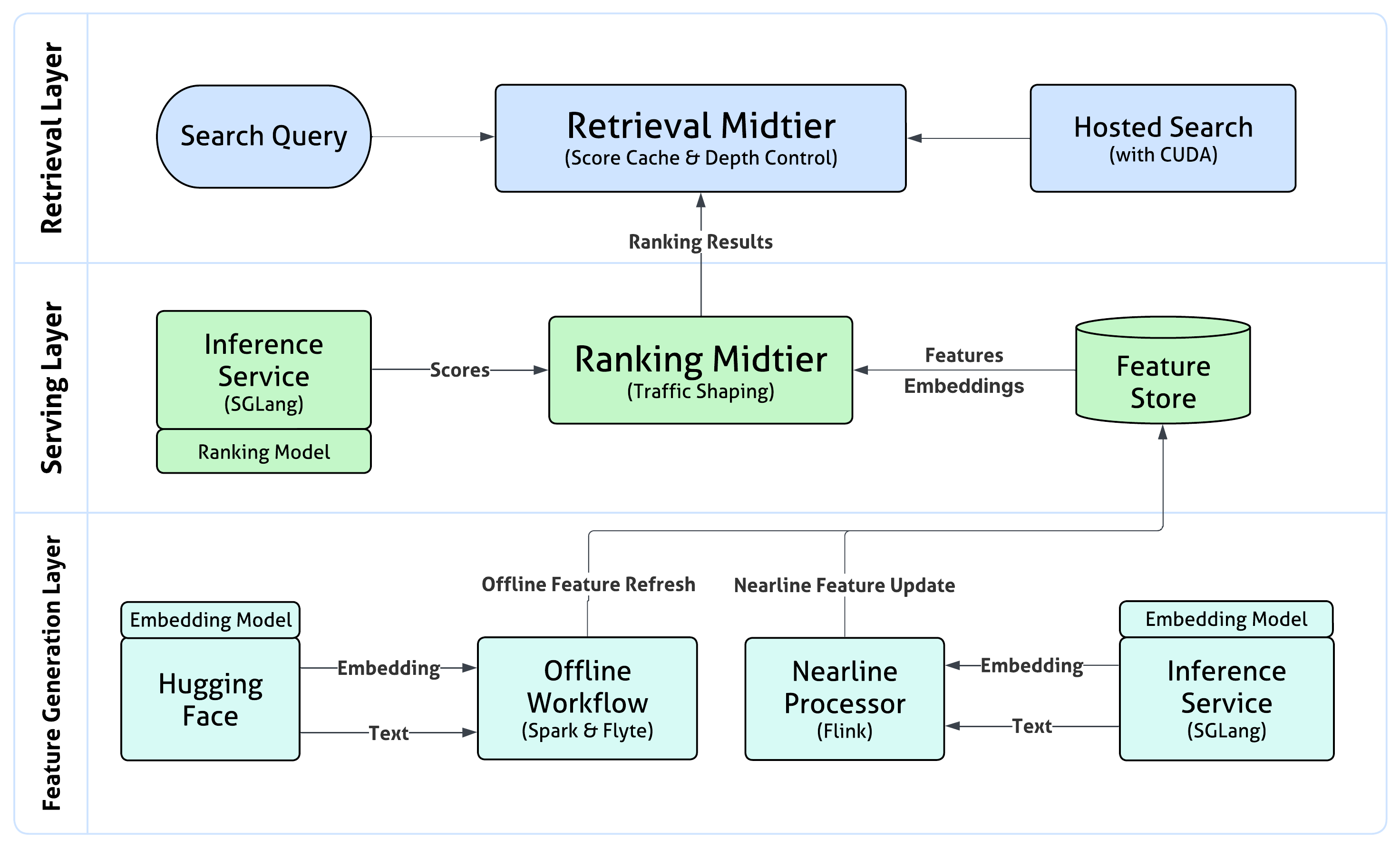}
    \captionsetup{font=small}
    \caption{Retrieval and ranking system architecture}
    \label{fig:system_diagram_fig}
\end{figure}

LLM-based ranking is dominated by long document context and prefill-heavy inference. To make
semantic search practical at scale, the production system
(Figure~\ref{fig:system_diagram_fig}) emphasizes end-to-end efficiency alongside model quality.
Pipeline controls such as score caching, ranking-depth control, and traffic shaping bound
worst-case cost, while hybrid offline/nearline feature computation precomputes expensive
signals for low-latency serving \cite{behdin2025scaling}. Additional compression techniques—
including structured pruning, offline document summarization, and mixed-input ranking (MixLM)
that replaces most item text with cached embedding tokens—further reduce online context and
improve GPU throughput \cite{li2025mixlmhighthroughputeffectivellm}. Together, these choices
enable high-throughput semantic ranking while preserving cross-encoder interaction quality.

\end{document}